\documentclass[12pt, fullpage]{article}
\usepackage{geometry}
\usepackage{graphicx} 
\usepackage{xcolor}
\usepackage{booktabs}
\usepackage{multirow}
\usepackage{enumitem}
\usepackage{hyperref}
\usepackage{natbib} 
\usepackage{array} 
\usepackage{placeins}
\usepackage{amsfonts}
\usepackage{amsmath}
\usepackage{amsthm}
\usepackage{algorithm}
\usepackage{algorithmic}

\newcommand{\bs}[1]{\boldsymbol{#1}}
\newcommand{\wh}[1]{\widehat{#1}}

\newcommand{\kk}{\kappa}
\hypersetup{hidelinks}
\newtheorem{theorem}{Theorem}

\newtheorem{proposition}{Proposition}

\begin{document}

\title{Semiparametric Receiver Operating Characteristic Analysis in the Presence of an Imperfect Reference Standard via a Box–Cox Density Ratio Model}
\author{Yi Chang, Siyan Liu, Qinglong Tian, and Pengfei Li}
\date{\small{Department of Statistics and Actuarial Science, University of Waterloo}}
\maketitle

\begin{abstract}
Receiver operating characteristic (ROC) analysis is commonly used to
evaluate the diagnostic accuracy of continuous biomarkers. In practice,
the true disease status may be unavailable and only a nominal disease
status provided by an imperfect reference standard is observed. Existing
nonparametric methods have been developed for ROC analysis in this
setting, but may suffer from reduced estimation efficiency, numerical
instability, or sensitivity to the choice of biomarker scale. We propose
a semiparametric method based on a Box--Cox density ratio model, which
links the biomarker distributions of the truly healthy and diseased
populations while leaving the baseline distribution unspecified. A key
feature of the proposed method is that the transformation parameter is
estimated from the data rather than specified in advance, allowing the
density-ratio structure to adapt to different transformation scales. We
develop an empirical likelihood approach for estimation and an
expectation-maximization algorithm for computation. We establish the
asymptotic distributions of estimators of the ROC curve, area under the
curve, Youden's index, and the sensitivity and specificity at the
Youden-optimal cutoff, and develop bootstrap confidence intervals and a
goodness-of-fit test. Simulation studies demonstrate that the proposed
method provides accurate and numerically stable estimation and inference
across a range of distributional settings without requiring the
transformation scale to be specified in advance. The proposed method is
illustrated using data from a malaria study.
\end{abstract}


\section{Introduction\label{bcdrm.sec1}}
Receiver operating characteristic (ROC) analysis is a widely used
statistical framework for evaluating the diagnostic accuracy of a
continuous biomarker in distinguishing diseased individuals from healthy
individuals \citep{pepe2003,zhou2026statistical}. Let $D\in\{0,1\}$
denote the true disease status, with $D=1$ indicating diseased and
$D=0$ indicating healthy, and let $F_0$ and $F_1$ denote the cumulative
distribution functions (cdfs) of the biomarker conditional on $D=0$
and $D=1$, respectively. When larger biomarker values are indicative
of disease, the ROC curve is defined as
\citep{yin2014joint,Hu2023Statistical}
\begin{equation}
\label{def.ROC}
\mathrm{ROC}(s)
=
1-F_1\{F_0^{-1}(1-s)\},
\qquad 0<s<1,
\end{equation}
where $s$ denotes the false-positive rate and $\mathrm{ROC}(s)$ is the
corresponding true-positive rate, or sensitivity. Important summary measures include the area under the ROC curve (AUC),
which characterizes the overall discriminatory ability of the biomarker,
and Youden's index, which measures the maximum separation between the
diseased and healthy populations. The cutoff attaining Youden's index,
together with the corresponding sensitivity and specificity, provides
clinically interpretable information for diagnostic decision-making. When the true
disease status $D$ is accurately observed, a large literature has
developed parametric, semiparametric, and nonparametric methods for
estimation and inference of the ROC curve and its associated summary
measures; see, for example,
\cite{pepe2003,nakas2024roc,zhou2026statistical} and references therein.

In practice, however, the true disease status $D$ may not be accurately
observed because a gold standard is unavailable or impractical to
implement due to cost, time, or ethical considerations. In such settings,
an imperfect reference standard is often used to provide a nominal disease
status \citep{Sun2024nonparametric}. For example, magnetic resonance
imaging (MRI) can be used in the diagnosis of brain tumors, although its
sensitivity and specificity are not perfect. Biopsy based on direct tissue
examination can provide a more definitive diagnosis, but its invasive
nature and associated risks limit its routine use \citep{taghipour2011}.
Thus, MRI may serve as a practically useful but imperfect reference
standard. Let $R\in\{0,1\}$ denote the nominal disease status determined
by the imperfect reference standard, with $R=1$ and $R=0$ indicating
classification as diseased and healthy, respectively. Because $R$ may
misclassify $D$, the nominally diseased group may contain truly healthy
individuals, while the nominally healthy group may contain truly diseased
individuals.

In this paper, we consider ROC curve analysis under a two-sample framework
in which one sample is drawn from the nominally healthy group ($R=0$)
and the other from the nominally diseased group ($R=1$). Define
\begin{equation}
\label{pi.def}
\pi_0=\Pr(D=0\mid R=0)
\quad\text{and}\quad
\pi_1=\Pr(D=1\mid R=1),
\end{equation}
which characterize the proportions of truly healthy and truly diseased
individuals in the two nominal groups, respectively. Under the assumption
that the biomarker and the reference standard are conditionally independent
given the true disease status $D$, the biomarker distributions in the
nominally healthy and nominally diseased groups are mixtures of the two
underlying true distributions. Specifically,
\begin{equation}
\label{def.Fstar}
F_0^*(t)=\pi_0F_0(t)+(1-\pi_0)F_1(t),
\qquad
F_1^*(t)=(1-\pi_1)F_0(t)+\pi_1F_1(t),
\end{equation}
where $F_0^*$ and $F_1^*$ denote the cdfs of the biomarker conditional
on $R=0$ and $R=1$, respectively. We assume that $\pi_0$ and $\pi_1$
are known, for example, from external validation studies. Such information
is necessary for identifying the underlying distributions $F_0$ and
$F_1$ from the observed mixture distributions $F_0^*$ and $F_1^*$
\citep{Sun2024nonparametric}. Thus, the primary statistical problem is
to recover $F_0$ and $F_1$ from the two contaminated samples and
subsequently estimate the ROC curve and its associated summary measures.

If the nominal disease status $R$ is naively treated as the true disease
status $D$, conventional ROC analysis is based on $F_0^*$ and $F_1^*$
rather than the target distributions $F_0$ and $F_1$. As a result, the
ROC curve and its associated summary measures can be substantially
underestimated \citep{Sun2024nonparametric}. To address
this issue, \citet{Sun2024nonparametric} proposed a fully nonparametric
approach that recovers $F_0$ and $F_1$ by inverting the mixture
relationships in \eqref{def.Fstar} using the empirical cdfs of the two
nominal groups, leading to consistent estimation of the ROC curve and
its associated summary measures. More recently,
\citet{Sun2025Likelihood} developed a likelihood-based nonparametric
(LBNP) approach that models the log-density ratio of $F_1$ to $F_0$
using B-splines while leaving the baseline distribution unspecified.
By incorporating smoothness of the density ratio within a likelihood
framework, LBNP can improve estimation efficiency over the fully
nonparametric approach while retaining substantial modeling flexibility.

Despite these improvements, several practical and inferential challenges
remain with the LBNP approach of
\citet{Sun2025Likelihood}. First, our numerical studies indicate that,
although the spline-based density-ratio model provides considerable
flexibility, its finite-sample performance can depend on the choice of
biomarker transformation. In practice, it may be unclear whether the
log-density ratio should be modeled on the original biomarker scale, the
logarithmic scale, or another transformed scale, and different choices
may lead to substantially different finite-sample performance. Second,
our numerical studies also reveal practical challenges associated with
the flexibility of the spline-based approach, including the selection
of the number of basis functions and the smoothing parameter, as well as
potential numerical instability in finite samples. Third, the available
theoretical results for the LBNP approach primarily establish consistency
of the estimated ROC curve and its associated summary measures, while
limiting distributions needed for statistical inference remain
unavailable. These considerations motivate introducing appropriate
structure into the relationship between the diseased and healthy
biomarker distributions to improve finite-sample performance and
computational stability and to facilitate statistical inference.

For the conventional setting in which the true disease status is
accurately observed, a substantial literature has developed parametric
and semiparametric methods for ROC analysis by imposing various degrees
of structure on the biomarker distributions or their relationship; see,
for example,
\cite{QinZhang2003,Yuan2021Semiparametric,bantis2024statistical,LJL2025,
zhou2026statistical} and references therein. Among these methods, the
Box--Cox transformation approach and the semiparametric density ratio
model (DRM) are particularly relevant to the development of our method.
The Box--Cox approach estimates a transformation parameter so that the
transformed biomarker can be modeled by normal distributions in the
healthy and diseased populations. It therefore allows the transformation
to be learned from the data, but relies on the assumed normality of the
transformed biomarker distributions. In contrast, the DRM models the
log-density ratio of $F_1$ to $F_0$ as a linear combination of
prespecified basis functions while leaving the baseline distribution
unspecified, thereby avoiding parametric specification of the individual
biomarker distributions. However, the basis functions in the standard
DRM are typically specified in advance. In practice, an appropriate
basis or transformation may instead be selected using the observed data.
Treating a data-adaptive transformation as if it had been prespecified ignores the associated uncertainty and may lead to confidence intervals with undercoverage, as demonstrated in our simulation study.

Motivated by these considerations, we propose a semiparametric model,
referred to as the Box--Cox density ratio model (BC-DRM), that combines
the data-adaptive transformation of the Box--Cox approach with the
distributional flexibility of the DRM. Specifically, we assume that
\begin{equation}
\label{def.bcdrm}
dF_1(t)
=
\exp\{\alpha+\beta B(t;\kappa)\}dF_0(t),
\end{equation}
where $\bs\theta=(\alpha,\beta,\kappa)^\top$ is the model parameter, $dF_0$ and $dF_1$ denote the probability measures associated with
$F_0$ and $F_1$, respectively, and
\[
B(t;\kappa)
=
\begin{cases}
(t^\kappa-1)/\kappa, & \kappa\neq 0,\\
\log(t), & \kappa=0,
\end{cases}
\qquad t>0,
\]
is the Box--Cox transformation. The baseline distribution $F_0$ is left completely
unspecified. The transformation parameter $\kappa$ is estimated jointly
with the other model parameters, allowing the transformation to be
learned from the data and, importantly, accounting for transformation
uncertainty in subsequent inference. The BC-DRM therefore avoids the
normality assumption required by the parametric Box--Cox approach and
the need to prespecify the basis functions in the standard DRM.
Moreover, despite its semiparametric nature, the BC-DRM encompasses
several commonly used pairs of distributions as special cases, including
two lognormal
distributions with a common variance on the log scale, two gamma
distributions with either a common shape or a common scale parameter,
and two Weibull distributions with a common shape parameter. At the same time, its finite-dimensional density-ratio structure provides
a parsimonious alternative to spline-based nonparametric modeling and a
convenient framework for stable computation and rigorous statistical
inference for ROC analysis in the presence of an imperfect reference
standard.

Our main contributions are summarized as follows.
\begin{itemize}

\item[(i)] We develop an empirical likelihood (EL) procedure for estimating
the BC-DRM parameters and the underlying distributions $F_0$ and $F_1$
from the two contaminated samples, and construct estimators of the ROC
curve, AUC, Youden's index, and the sensitivity and specificity at the
Youden-optimal cutoff.

\item[(ii)] We establish asymptotic normality for the proposed estimators of the ROC curve and its associated summary measures and develop corresponding statistical inference that accounts for uncertainty in estimating the Box–Cox transformation parameter. We also develop a
goodness-of-fit procedure for assessing the adequacy of the proposed
BC-DRM.
\item[(iii)] We develop an expectation-maximization (EM) algorithm to
facilitate computation in the presence of the unobserved true disease
status.

\item[(iv)] Extensive simulation studies demonstrate favorable finite-sample
performance and numerical stability of the proposed method relative to
existing approaches, satisfactory finite-sample size and power of the proposed
goodness-of-fit test, and the importance of accounting for transformation
uncertainty in statistical inference.

\item[(v)] An application to malaria data illustrates the practical utility
of the proposed method for ROC analysis in the presence of an imperfect
reference standard.

\end{itemize}

The remainder of the paper is organized as follows.
Section~\ref{bcdrm.sec2} presents the problem setup, establishes identifiability under
the BC-DRM, develops an empirical likelihood procedure for estimating
the model parameters and the underlying biomarker distributions, and
constructs estimators of the ROC curve and its associated summary
measures. Section~\ref{bcdrm.sec3} establishes the asymptotic properties of the proposed
estimators, develops statistical inference for the ROC curve and its
associated summary measures, and presents a goodness-of-fit procedure
for assessing the adequacy of the proposed model. Section~\ref{bcdrm.sec4} develops an
EM algorithm for numerical computation. Section~\ref{bcdrm.sec5} reports simulation
studies evaluating the finite-sample performance of the proposed method,
and Section~\ref{bcdrm.sec6} illustrates the method using data from a malaria study.
Section~\ref{bcdrm.sec7} concludes the paper with a discussion of limitations and directions for
future research.

\section{Problem setup and estimation\label{bcdrm.sec2}}

\subsection{Problem setup and identifiability}

Let $T$ denote the positive continuous biomarker of interest. Recall that
$F_0$ and $F_1$ are the conditional cdfs of $T$ given $D=0$ and $D=1$,
respectively. We formally state below the two assumptions described in
the Introduction:
\begin{enumerate}
    \item[\textnormal{A1.}] The biomarker $T$ and the nominal disease
    status $R$ are conditionally independent given the true disease
    status $D$, i.e., $T\perp R\mid D$;

    \item[\textnormal{A2.}] $\pi_0$ and $\pi_1$ defined in
    \eqref{pi.def} are known and satisfy $\pi_0+\pi_1>1$.
\end{enumerate}

We observe two independent samples
\[
X_{1},\ldots,X_{n_0}\sim 
F_0^*(t)=\pi_0F_0(t)+(1-\pi_0)F_1(t);
~~
Y_{1},\ldots,Y_{n_1}\sim 
F_1^*(t)=(1-\pi_1)F_0(t)+\pi_1F_1(t),
\]
from the nominally healthy and nominally diseased groups, 
respectively, where the mixture representations are given in 
\eqref{def.Fstar}.

We further assume that $F_0$ and $F_1$ satisfy the BC-DRM in
\eqref{def.bcdrm}. The unknown quantities are the finite-dimensional 
parameter ${\bs\theta}=(\alpha,\beta,\kk)^\top$ and the baseline 
distribution $F_0$. Before developing the estimation procedure, we first
establish their identifiability. The following proposition gives the
result.

\begin{proposition}
\label{prop1}
Suppose that Assumptions A1 and A2 hold and that $\beta\neq0$. Then
${\bs\theta}$ and $F_0$ are identifiable.
\end{proposition}

The proof of Proposition~\ref{prop1} is provided in Section~1 of the
Supplementary Material. The condition $\beta\neq0$ corresponds to the nondegenerate setting in
which $F_0\neq F_1$, so that the biomarker distributions differ between
the truly diseased and healthy populations.
When $\beta=0$, we have $F_0=F_1$, and the transformation parameter
$\kappa$ is not identifiable.

\subsection{Empirical likelihood estimation}

Under the mixture structure in \eqref{def.Fstar} and the BC-DRM in
\eqref{def.bcdrm}, the observed-data likelihood can be written as
\begin{align}
\prod_{i=1}^{n_0}dF_0^*(X_i)
\prod_{j=1}^{n_1}dF_1^*(Y_j)
={}&
\prod_{i=1}^{n_0}
\left[\pi_0+(1-\pi_0)
\exp\{\alpha+\beta B(X_i;\kk)\}\right]dF_0(X_i)\nonumber\\
&\times
\prod_{j=1}^{n_1}
\left[1-\pi_1+\pi_1
\exp\{\alpha+\beta B(Y_j;\kk)\}\right]dF_0(Y_j).
\label{full.like}
\end{align}

Let $n=n_0+n_1$ and
\(
{\bs T}
=
(T_1,\ldots,T_n)^\top
=
(X_1,\ldots,X_{n_0},Y_1,\ldots,Y_{n_1})^\top
\)
denote the pooled biomarker sample. Following the empirical likelihood
(EL) framework of \citet{owen2001empirical}, we place probability mass
\(
p_i=dF_0(T_i)
\)
at each pooled observation $T_i$, $i=1,\ldots,n$.

Let $\bs p=(p_1,\ldots,p_n)^\top$. Substituting
$p_i=dF_0(T_i)$ into \eqref{full.like} and taking logarithms, the
log-EL function for $(\bs\theta,\bs p)$ is
\begin{align*}
\ell(\bs\theta,\bs p)
={}&
\sum_{i=1}^{n}\log p_i
+\sum_{i=1}^{n_0}
\log\left[\pi_0+(1-\pi_0)
\exp\{\alpha+\beta B(X_i;\kk)\}\right]\nonumber\\
&+
\sum_{j=1}^{n_1}
\log\left[1-\pi_1+\pi_1
\exp\{\alpha+\beta B(Y_j;\kk)\}\right].
\label{log.el}
\end{align*}
The probability masses are subject to
\begin{equation}
p_i>0,\qquad
\sum_{i=1}^n p_i=1,\qquad
\sum_{i=1}^n p_i
\left[\exp\{\alpha+\beta B(T_i;\kk)\}-1\right]=0.
\label{eq:el_constraint}
\end{equation}
The first two constraints ensure that the probability masses define a
valid cdf $F_0$, while the third, together with \eqref{def.bcdrm},
ensures that the corresponding $F_1$ is also a valid cdf.

The maximum empirical likelihood estimator (MELE) of
$(\bs\theta,\bs p)$ is defined as
\begin{equation}
\label{MELE}
(\wh{\bs\theta},\wh{\bs p})
=
\arg\max_{\bs\theta,\,\bs p}
\ell(\bs\theta,\bs p)
\quad\mbox{subject to }\eqref{eq:el_constraint}.
\end{equation}
In general, the MELE $(\wh{\bs\theta},\wh{\bs p})$ does not admit a
closed-form solution. In Section~\ref{bcdrm.sec4}, we develop an EM
algorithm for its numerical computation.

Given $(\wh{\bs\theta},\wh{\bs p})$, the cdfs $F_0$ and $F_1$ are 
estimated, respectively, by
\begin{equation}
\label{F.est}
\wh F_0(t)
=
\sum_{i=1}^n\wh p_i I(T_i\leq t),
\qquad
\wh F_1(t)
=
\sum_{i=1}^n\wh p_i
\exp\{\wh\alpha+\wh\beta B(T_i;\wh\kk)\}
I(T_i\leq t),
\end{equation}
where $I(\cdot)$ denotes the indicator function.
These estimators form the basis for estimating the ROC curve and its
associated summary measures in the next subsection.

\subsection{Estimation of the ROC curve and summary measures}

Recall the definition of the ROC curve in \eqref{def.ROC}. In addition
to the ROC curve, we consider several commonly used summary measures of
diagnostic accuracy. The AUC is defined as
\[
\mathrm{AUC}
=
\int_0^1\mathrm{ROC}(s)\,ds
=
\int F_0(t)\,dF_1(t).
\]
Youden's index is defined as
\[
J
=
\sup_t\{F_0(t)-F_1(t)\}
=
F_0(c_0)-F_1(c_0),
\]
where the Youden-optimal cutoff $c_0$ is
\[
c_0
=
\arg\max_t\{F_0(t)-F_1(t)\}.
\]
The sensitivity and specificity at $c_0$ are given by
\[
\eta=1-F_1(c_0),
\qquad
\tau=F_0(c_0),
\]
respectively.

Based on $\wh F_0$ and $\wh F_1$ in \eqref{F.est}, the ROC curve and AUC
are estimated by the plug-in estimators
\begin{equation*}
\label{ROC.AUC.est}
\widehat{\mathrm{ROC}}(s)
=
1-\wh F_1\{\wh F_0^{-1}(1-s)\},
\quad 0<s<1,
\qquad
\widehat{\mathrm{AUC}}
=
\int \wh F_0(t)\,d\wh F_1(t).
\end{equation*}

The Youden-optimal cutoff can be estimated more directly by exploiting
the structure of the BC-DRM. Under \eqref{def.bcdrm}, $c_0$ is
characterized by
\[
\alpha+\beta B(c_0;\kappa)=0.
\]
Hence, we estimate $c_0$ by solving
\[
\wh\alpha+\wh\beta B(\wh c_0;\wh\kk)=0,
\]
which gives
\begin{equation*}
\wh c_0
=
B^{-1}\left(-\frac{\wh\alpha}{\wh\beta};\wh\kk\right).
\end{equation*}
It is shown in Section~2 of the Supplementary Material that, if
$\wh\beta\neq0$, $\wh c_0$ exists and is unique in
$(T_{(1)},T_{(n)})$, where
$T_{(1)}=\min_{1\leq i\leq n}T_i$ and
$T_{(n)}=\max_{1\leq i\leq n}T_i$.

Using $\wh c_0$, we estimate $(J,\eta,\tau)$ by
\begin{equation*}
\label{J.eta.tau.est}
\wh J
=
\wh F_0(\wh c_0)-\wh F_1(\wh c_0),
\qquad
\wh\eta
=
1-\wh F_1(\wh c_0),
\qquad
\wh\tau
=
\wh F_0(\wh c_0).
\end{equation*}

\section{Statistical inference and goodness-of-fit\label{bcdrm.sec3}}

\subsection{Asymptotic properties}

In this subsection, we establish the asymptotic properties of the
estimated ROC curve and its summary measures. Let $\mathrm{ROC}_0(s)$,
$\mathrm{AUC}_0$, $J_0$, $\eta_0$, and $\tau_0$ denote the true values
of the ROC curve, AUC, Youden's index, and the sensitivity and specificity
at the Youden-optimal cutoff, respectively. The technical regularity
conditions and proofs of the results below are provided in Section~3 of
the Supplementary Material. We use $\stackrel{d}{\longrightarrow}$ to denote convergence in distribution.

\begin{theorem}
\label{thm.asymp}
Suppose that Assumptions A1 and A2 hold. Under the regularity conditions
given in Section~3 of the Supplementary Material, the following results
hold:
\begin{enumerate}
    \item[(i)] For any fixed $s\in(0,1)$,
    \[
    \sqrt{n}\{\widehat{\mathrm{ROC}}(s)-\mathrm{ROC}_0(s)\}
    \stackrel{d}{\longrightarrow}
    N\{0,\sigma_{\mathrm{ROC}}^2(s)\}.
    \]

    \item[(ii)]
    \[
    \sqrt{n}(\widehat{\mathrm{AUC}}-\mathrm{AUC}_0)
    \stackrel{d}{\longrightarrow}
    N(0,\sigma_{\mathrm{AUC}}^2).
    \]

    \item[(iii)]
    \[
    \sqrt{n}(\wh J-J_0)
    \stackrel{d}{\longrightarrow}
    N(0,\sigma_J^2).
    \]

    \item[(iv)]
    \[
    \sqrt{n}(\wh\eta-\eta_0)
    \stackrel{d}{\longrightarrow}
    N(0,\sigma_\eta^2).
    \]

    \item[(v)]
    \[
    \sqrt{n}(\wh\tau-\tau_0)
    \stackrel{d}{\longrightarrow}
    N(0,\sigma_\tau^2).
    \]
\end{enumerate}
The explicit forms of $\sigma_{\mathrm{ROC}}^2(s)$,
$\sigma_{\mathrm{AUC}}^2$, $\sigma_J^2$, $\sigma_\eta^2$, and
$\sigma_\tau^2$ are provided in Section~3 of the Supplementary Material.
\end{theorem}

Theorem~\ref{thm.asymp} shows that the proposed estimator of the ROC
curve at each fixed $s\in(0,1)$ and the estimators of its summary
measures are $\sqrt{n}$-consistent and asymptotically normal. For ROC
analysis with an imperfect reference standard, asymptotic normality has
previously been established for estimators of the ROC curve and AUC
under the fully nonparametric approach of
\citet{Sun2024nonparametric}, whereas corresponding results are not
available for the likelihood-based nonparametric approach of
\citet{Sun2025Likelihood}. To the best of our knowledge, the asymptotic
normality results for the estimators of Youden's index and the
sensitivity and specificity at the Youden-optimal cutoff are new in
the literature on ROC analysis with an imperfect reference standard.

\subsection{Bootstrap confidence intervals}

Although Theorem~\ref{thm.asymp} provides a theoretical basis for
statistical inference, the corresponding asymptotic variance expressions
generally have complicated forms, making direct variance estimation
cumbersome in practice. We therefore use the nonparametric bootstrap to
construct percentile bootstrap confidence intervals (CIs) for the ROC
curve and its summary measures. We describe the procedure for
$\mathrm{ROC}(s)$ at a fixed $s\in(0,1)$; the procedures for AUC, $J$,
$\eta$, and $\tau$ are analogous.

For $b=1,\ldots,M$, the bootstrap procedure consists of the following
steps:
\begin{enumerate}
    \item Independently draw, with replacement, a sample
    $\{X_i^{(b)}\}_{i=1}^{n_0}$ from $\{X_i\}_{i=1}^{n_0}$ and a sample
    $\{Y_j^{(b)}\}_{j=1}^{n_1}$ from $\{Y_j\}_{j=1}^{n_1}$.

    \item Apply the proposed estimation procedure to the two bootstrap
    samples to obtain $\widehat{\mathrm{ROC}}^{(b)}(s)$.
\end{enumerate}

Let $q_{\gamma,M}$ denote the empirical $\gamma$-quantile of the $M$
bootstrap estimates
$\{\widehat{\mathrm{ROC}}^{(b)}(s):b=1,\ldots,M\}$. A
$100(1-\delta)\%$ percentile bootstrap CI for $\mathrm{ROC}(s)$ is
then given by
\[
\left[q_{\delta/2,M},q_{1-\delta/2,M}\right].
\]
Percentile bootstrap CIs for AUC, $J$, $\eta$, and $\tau$ are constructed
analogously using their corresponding bootstrap estimates.

We note that the transformation parameter $\kappa$ is re-estimated in
each bootstrap sample, so that the resulting CIs account for its
estimation uncertainty. In Section~5 of the Supplementary Material, we
provide numerical evidence showing that fixing $\kappa$ in the bootstrap
procedure can lead to substantial undercoverage of the resulting CIs.

\subsection{Goodness-of-fit test\label{bcdrm.main.sec3.3}}

The proposed BC-DRM imposes a specific structure on the relationship
between $F_0$ and $F_1$ through \eqref{def.bcdrm}. It is therefore
important to assess whether the model provides an adequate fit to the
observed data. We develop a goodness-of-fit (GOF) test for the BC-DRM.

Let $\widetilde F_0^*$ and $\widetilde F_1^*$ denote the empirical cdfs
of the nominally healthy and nominally diseased samples, respectively:
\[
\widetilde F_0^*(t)
=
\frac{1}{n_0}\sum_{i=1}^{n_0}I(X_i\leq t),
\qquad
\widetilde F_1^*(t)
=
\frac{1}{n_1}\sum_{j=1}^{n_1}I(Y_j\leq t).
\]
Under the BC-DRM, the fitted cdfs $F_0^*$ and $F_1^*$ are obtained by
substituting $\wh F_0$ and $\wh F_1$ in \eqref{F.est} into the mixture
representations in \eqref{def.Fstar}. Specifically,
\[
\wh F_0^*(t)
=
\pi_0\wh F_0(t)+(1-\pi_0)\wh F_1(t),
\qquad
\wh F_1^*(t)
=
(1-\pi_1)\wh F_0(t)+\pi_1\wh F_1(t).
\]

To assess the goodness of fit of the BC-DRM, we consider the supremum
distances between the empirical and fitted cdfs in the two nominal
groups. Specifically, define
\[
D_{n,0}
=
\sup_t\left|
\widetilde F_0^*(t)-\wh F_0^*(t)
\right|,
\qquad
D_{n,1}
=
\sup_t\left|
\widetilde F_1^*(t)-\wh F_1^*(t)
\right|.
\]
Following \citet{Zhang2002}, we use
\[
D_n
=
\frac{n_0}{n}D_{n,0}
+
\frac{n_1}{n}D_{n,1}
\]
as the GOF test statistic, with large values of $D_n$ providing evidence
against the BC-DRM. We also considered alternative ways of combining
$D_{n,0}$ and $D_{n,1}$ and found that the above weighted statistic
generally performed similarly but exhibited slightly better power in
some simulation settings. We therefore use $D_n$ throughout.

The null distribution of $D_n$ is not readily available in a tractable form. We therefore use a model-based bootstrap
procedure to approximate its null distribution and compute the
corresponding $p$-value. Specifically, for $b=1,\ldots,M$, we carry out
the following steps:
\begin{enumerate}
    \item Generate a bootstrap sample
    $\{X_i^{(b)}\}_{i=1}^{n_0}$ from $\wh F_0^*$ and, independently, a
    bootstrap sample $\{Y_j^{(b)}\}_{j=1}^{n_1}$ from $\wh F_1^*$.

    \item Apply the proposed estimation procedure to the two bootstrap
    samples to obtain the fitted cdfs $\wh F_0^{*(b)}$ and
    $\wh F_1^{*(b)}$.

    \item Let $\widetilde F_0^{*(b)}$ and $\widetilde F_1^{*(b)}$
    denote the empirical cdfs of the bootstrap samples
    $\{X_i^{(b)}\}_{i=1}^{n_0}$ and
    $\{Y_j^{(b)}\}_{j=1}^{n_1}$, respectively, and compute
    \[
    D_{n,0}^{(b)}
    =
    \sup_t
    \left|
    \widetilde F_0^{*(b)}(t)-\wh F_0^{*(b)}(t)
    \right|,
    \qquad
    D_{n,1}^{(b)}
    =
    \sup_t
    \left|
    \widetilde F_1^{*(b)}(t)-\wh F_1^{*(b)}(t)
    \right|.
    \]
Set
\[
D_n^{(b)}
=
\frac{n_0}{n}D_{n,0}^{(b)}
+
\frac{n_1}{n}D_{n,1}^{(b)}.
\]
\end{enumerate}

The bootstrap $p$-value is then estimated by
\[
\widehat p
=
\frac{1}{M}
\sum_{b=1}^M
I\{D_n^{(b)}\geq D_n\}.
\]
For a test with significance level $\delta$, the BC-DRM is rejected if
$\widehat p\leq\delta$. In Section~5 of the Supplementary Material, we
provide numerical evidence supporting the performance of the proposed
GOF test. 

For both bootstrap procedures described in this section, we
use $M=500$ throughout the simulation studies and real-data analysis.

\section{EM algorithm\label{bcdrm.sec4}}

In this section, we develop an EM algorithm \citep{dempster1977maximum} to compute the MELE
$(\wh{\bs\theta},\wh{\bs p})$ defined in \eqref{MELE}.

For $i=1,\ldots,n_0$, let $D_i$ denote the latent true disease status
of the $i$th individual in the $R=0$ group, and for
$j=1,\ldots,n_1$, let $D_{n_0+j}$ denote the latent true disease status
of the $j$th individual in the $R=1$ group.

Given the latent true disease statuses, the complete-data log-likelihood
is
\begin{align*}
\ell_c(\bs\theta,\bs p)
={}&
\sum_{i=1}^{n_0}
\Big[
(1-D_i)\log\pi_0
+D_i\log(1-\pi_0)
+D_i\{\alpha+\beta B(X_i;\kk)\}
\Big]
\nonumber\\
&+
\sum_{j=1}^{n_1}
\Big[
(1-D_{n_0+j})\log(1-\pi_1)
+D_{n_0+j}\log\pi_1
+D_{n_0+j}\{\alpha+\beta B(Y_j;\kk)\}
\Big]
\nonumber\\
&+
\sum_{k=1}^n\log p_k.
\label{complete.loglike}
\end{align*}

The EM algorithm iterates between an E-step and an M-step. Let
$(\bs\theta^{(r)},\bs p^{(r)})$ denote the parameter values after
the $r$th iteration, with
$(\bs\theta^{(0)},\bs p^{(0)})$ denoting the initial values. In the
E-step, we compute the conditional expectations of the latent true
disease statuses given the observed data and the current parameter
values. In the M-step, we update $(\bs\theta,\bs p)$ by maximizing the
resulting conditional expectation of the complete-data log-likelihood
subject to the constraints in \eqref{eq:el_constraint}.

\textbf{E-step.}
For $i=1,\ldots,n_0$, we compute
\begin{equation}
w_{i0}^{(r+1)}
=
E\{D_i\mid{\bs T};
\bs\theta^{(r)},\bs p^{(r)}\}
=
\frac{(1-\pi_0)
\exp\{\alpha^{(r)}
+\beta^{(r)}B(X_i;\kk^{(r)})\}}
{\pi_0+(1-\pi_0)
\exp\{\alpha^{(r)}
+\beta^{(r)}B(X_i;\kk^{(r)})\}},
\label{E.X}
\end{equation}
and, for $j=1,\ldots,n_1$,
\begin{align}
w_{j1}^{(r+1)}
=
E\{D_{n_0+j}\mid{\bs T};
\bs\theta^{(r)},\bs p^{(r)}\}
=
\frac{\pi_1
\exp\{\alpha^{(r)}
+\beta^{(r)}B(Y_j;\kk^{(r)})\}}
{1-\pi_1+\pi_1
\exp\{\alpha^{(r)}
+\beta^{(r)}B(Y_j;\kk^{(r)})\}}.
\label{E.Y}
\end{align}
It follows that the conditional expectation of the complete-data
log-likelihood, given ${\bs T}$ and the current parameter values, is
\begin{align*}
Q^{(r)}(\bs\theta,\bs p)
&=
E\left\{
\ell_c(\bs\theta,\bs p)
\mid {\bs T};
\bs\theta^{(r)},\bs p^{(r)}
\right\}
\nonumber\\
&=
\sum_{i=1}^{n_0}
w_{i0}^{(r+1)}
\{\alpha+\beta B(X_i;\kk)\}
+
\sum_{j=1}^{n_1}
w_{j1}^{(r+1)}
\{\alpha+\beta B(Y_j;\kk)\}
\nonumber\\
&\quad+
\sum_{k=1}^n\log p_k+C^{(r)},
\label{Q.function}
\end{align*}
where $C^{(r)}$ does not depend on $(\bs\theta,\bs p)$.

\textbf{M-step.}
We update $(\bs\theta^{(r)},\bs p^{(r)})$ to
$(\bs\theta^{(r+1)},\bs p^{(r+1)})$ by maximizing
$Q^{(r)}(\bs\theta,\bs p)$ subject to \eqref{eq:el_constraint}. For this constrained maximization, let
\[
\lambda^{(r+1)}
=
\frac1n\left\{\sum_{i=1}^{n_0}w_{i0}^{(r+1)}
+
\sum_{j=1}^{n_1}w_{j1}^{(r+1)}\right\}
\]
and define
\[
\alpha^*
=
\alpha+
\log\left\{
\frac{\lambda^{(r+1)}}{1-\lambda^{(r+1)}}
\right\}.
\]
We show in Section~4 of the Supplementary Material that
\begin{equation*}
\label{theta.update}
\left(
\alpha^{*(r+1)},\beta^{(r+1)},\kk^{(r+1)}
\right)
=
\arg\max_{\alpha^*,\beta,\kk}
H^{(r)}(\alpha^*,\beta,\kk),
\end{equation*}
where
\begin{align}
H^{(r)}(\alpha^*,\beta,\kk)
={}&
\sum_{i=1}^{n_0}
w_{i0}^{(r+1)}
\{\alpha^*+\beta B(X_i;\kk)\}
+
\sum_{j=1}^{n_1}
w_{j1}^{(r+1)}
\{\alpha^*+\beta B(Y_j;\kk)\}
\nonumber\\
&-
\sum_{k=1}^{n}
\log\left[
1+\exp\{\alpha^*+\beta B(T_k;\kk)\}
\right].
\label{M.objective}
\end{align}

The maximization of $H^{(r)}(\alpha^*,\beta,\kk)$ can be carried out
in two steps. First, for a fixed value of $\kk$,
$H^{(r)}(\alpha^*,\beta,\kk)$ has the form of a weighted logistic
log-likelihood. Hence, the maximizers $(\alpha^*,\beta)$ can be obtained
by fitting a weighted logistic regression, for example, using the R
function \texttt{glmnet}. Substituting these maximizers into
$H^{(r)}(\alpha^*,\beta,\kk)$ yields a one-dimensional profile objective
function for $\kk$. We then maximize this profile objective over $\kk$
using a one-dimensional numerical optimization procedure, such as the
R function \texttt{optimize}. The resulting maximizers are denoted by
$(\alpha^{*(r+1)},\beta^{(r+1)},\kk^{(r+1)})$.

Once $(\alpha^{*(r+1)},\beta^{(r+1)},\kk^{(r+1)})$ is obtained, we set
\begin{equation}
\label{alpha.update}
\alpha^{(r+1)}
=
\alpha^{*(r+1)}
-
\log\left\{
\frac{\lambda^{(r+1)}}{1-\lambda^{(r+1)}}
\right\},
\end{equation}
and update the probability masses by
\begin{equation}
\label{p.update}
p_k^{(r+1)}
=
\frac1n 
\frac{1}
{1+
\lambda^{(r+1)}
\left[
\exp\{\alpha^{(r+1)}
+\beta^{(r+1)}B(T_k;\kk^{(r+1)})\}-1
\right]},
\qquad k=1,\ldots,n.
\end{equation}

We iterate the E-step and M-step until convergence. Specifically, the
algorithm is terminated when the change in the log-EL function between
two consecutive iterations is less than a prespecified tolerance
$\epsilon$, that is,
\[
\left|
\ell(\bs\theta^{(r+1)},\bs p^{(r+1)})
-
\ell(\bs\theta^{(r)},\bs p^{(r)})
\right|
<\epsilon.
\]
We summarize the above procedure in Algorithm~\ref{alg:EM}.

\begin{algorithm}[!h]
\caption{EM algorithm for computing the MELE}
\label{alg:EM}
\begin{algorithmic}[1]
\REQUIRE Nominally healthy sample $\{X_i\}_{i=1}^{n_0}$;
nominally diseased sample $\{Y_j\}_{j=1}^{n_1}$; known
$\pi_0$ and $\pi_1$; initial values
$(\bs\theta^{(0)},\bs p^{(0)})$; prespecified tolerance $\epsilon$.

\STATE Set $r=0$.

\STATE \textbf{E-step:} Compute $w_{i0}^{(r+1)}$,
$i=1,\ldots,n_0$, and $w_{j1}^{(r+1)}$,
$j=1,\ldots,n_1$, according to \eqref{E.X} and \eqref{E.Y}.

\STATE \textbf{M-step:} Obtain
$(\alpha^{*(r+1)},\beta^{(r+1)},\kk^{(r+1)})$ by maximizing
\eqref{M.objective} as described above. Then update
$\alpha^{(r+1)}$ and $\bs p^{(r+1)}$ according to
\eqref{alpha.update} and \eqref{p.update}, respectively.

\STATE If
\[
\left|
\ell(\bs\theta^{(r+1)},\bs p^{(r+1)})
-
\ell(\bs\theta^{(r)},\bs p^{(r)})
\right|
<\epsilon,
\]
terminate the algorithm. Otherwise, set $r\leftarrow r+1$ and repeat
the E-step and M-step.

\end{algorithmic}
\end{algorithm}

The following theorem establishes the monotonicity of the proposed
EM algorithm.

\begin{theorem}
\label{thm.EM}
Let $(\bs\theta^{(r)},\bs p^{(r)})$ denote the parameter values after
the $r$th iteration of Algorithm~\ref{alg:EM}. Then, for $r\geq 0$,
\[
\ell\{\bs\theta^{(r+1)},\bs p^{(r+1)}\}
\geq
\ell\{\bs\theta^{(r)},\bs p^{(r)}\}.
\]
\end{theorem}

The proof of Theorem~\ref{thm.EM} is provided in Section~4 of the
Supplementary Material. The theorem shows that the log-EL function is
nondecreasing along the iterations of Algorithm~\ref{alg:EM}. Since the
log-EL function is bounded above, the sequence of log-EL values
converges. This result, however, does not guarantee convergence to the
global maximum. In practice, we therefore run Algorithm~\ref{alg:EM}
from multiple initial values and use the solution that yields the
largest log-EL value.
The tolerance $\epsilon$ in Algorithm~\ref{alg:EM} controls the
convergence criterion. Throughout the numerical studies, we set
$\epsilon=10^{-6}$.

\section{Simulation studies\label{bcdrm.sec5}}

\subsection{Simulation settings}

In this subsection, we compare the performance of the proposed method
(denoted by ``Our’’) with the following competing methods:
\begin{itemize}
\item \textbf{NP:} the fully nonparametric method of
\citet{Sun2024nonparametric};

\item \textbf{LBNP:} the likelihood-based nonparametric method of 
\citet{Sun2025Likelihood}, applied to the original biomarker scale;
\item \textbf{LBNP-log:} the likelihood-based nonparametric method 
of \citet{Sun2025Likelihood}, applied after log transformation of 
the biomarker.

\end{itemize}
The default implementation of LBNP uses 50 B-spline basis functions,
which resulted in frequent fitting failures in our simulations. We
therefore use 10 basis functions, the smallest number considered in
\citet{Sun2025Likelihood}. 
The simulation studies in Section~S4 of their Supplementary Material show that using 10 basis functions yields performance comparable to that using 50 basis functions.
All results for LBNP and
LBNP-log reported below are based on 10 basis functions, and fitting
failures are reported together with the estimation results.

We consider three distributional scenarios under which the BC-DRM
\eqref{def.bcdrm} holds with different values of the transformation
parameter $\kk$:
\begin{enumerate}
    \item[(1)] $F_0=\mathrm{Lognormal}(a_0,b_0=1)$ and
    $F_1=\mathrm{Lognormal}(a_1,b_1=1)$, corresponding to $\kk=0$;

    \item[(2)] $F_0=\mathrm{Weibull}(a_0,b_0=1/2)$ and
    $F_1=\mathrm{Weibull}(a_1,b_1=1/2)$, corresponding to $\kk=1/2$;

    \item[(3)] $F_0=\mathrm{Gamma}(a_0=1,b_0)$ and
    $F_1=\mathrm{Gamma}(a_1=1,b_1)$, corresponding to $\kk=1$.
\end{enumerate}
Here, $\mathrm{Lognormal}(a,b)$ denotes a lognormal distribution such
that $\log(T)$ has mean $a$ and variance $b$; $\mathrm{Weibull}(a,b)$ denotes a Weibull distribution
with scale $a$ and shape $b$; and 
$\mathrm{Gamma}(a,b)$ denotes a gamma distribution with shape $a$ and
scale $b$. These scenarios allow us to evaluate the
performance of the proposed method under different transformations
within the BC-DRM family.

For each scenario, we fix $F_0$ and choose the parameter of $F_1$ such
that $J_0\in\{0.3,0.5\}$, representing two levels of separation between
the truly healthy and diseased populations. We consider
$\pi_0=\pi_1\in\{0.75,0.90\}$ and
$n_0=n_1\in\{100,300,500\}$. Each simulation setting is repeated
$N=1000$ times. Table~\ref{tab:truth} reports the distributional
parameters and the corresponding true values of AUC, $J_0$, $\eta_0$,
and $\tau_0$. Since the results are qualitatively similar for the two
levels of classification accuracy, we report those for
$\pi_0=\pi_1=0.90$ in the main text and defer the results for
$\pi_0=\pi_1=0.75$ to the Supplementary Material.

\begin{table}[!htbp]
\centering
\caption{Parameter configurations and true diagnostic accuracy measures
for the three simulation scenarios.}
\label{tab:truth}
\begin{tabular}{lcccccccc}
\toprule
Distribution & $\mathrm{AUC}_0$ & $J_0$ & $\eta_0$ & $\tau_0$
& $a_0$ & $b_0$ & $a_1$ & $b_1$ \\
\midrule
Lognormal $(\kk=0)$
& 0.71 & 0.30 & 0.65 & 0.65 & 0.00 & 1.00 & 0.77 & 1.00 \\
Lognormal $(\kk=0)$
& 0.83 & 0.50 & 0.75 & 0.75 & 0.00 & 1.00 & 1.35 & 1.00 \\
\midrule
Weibull $(\kk=1/2)$
& 0.70 & 0.30 & 0.53 & 0.77 & 0.50 & 0.50 & 2.68 & 0.50 \\
Weibull $(\kk=1/2)$
& 0.82 & 0.50 & 0.65 & 0.85 & 0.50 & 0.50 & 9.73 & 0.50 \\
\midrule
Gamma $(\kk=1)$
& 0.70 & 0.30 & 0.53  & 0.77 & 1.00 & 1.00 & 1.00 & 2.31 \\
Gamma $(\kk=1)$
& 0.81 & 0.50 & 0.65 & 0.85 & 1.00 & 1.00 & 1.00 & 4.40 \\
\bottomrule
\end{tabular}
\end{table}

\subsection{Evaluation of point estimation}

In this subsection, we assess point estimation performance using
percentage relative bias (\%RB) and mean squared error (MSE). Let
$\widehat{\mathrm{AUC}}^{(i)}$, $i=1,\ldots,N$, denote the AUC estimate
from the $i$th simulation replicate. We define
\[
\mathrm{\%RB}(\mathrm{AUC})
=
\frac{1}{N}
\sum_{i=1}^{N}
\frac{\widehat{\mathrm{AUC}}^{(i)}-\mathrm{AUC}_0}
{\mathrm{AUC}_0}
\times 100\%,
\qquad
\mathrm{MSE}(\mathrm{AUC})
=
\frac{1}{N}
\sum_{i=1}^{N}
\left\{
\widehat{\mathrm{AUC}}^{(i)}-\mathrm{AUC}_0
\right\}^{2},
\]
where $\mathrm{AUC}_0$ denotes the true AUC. The corresponding criteria
for $\mathrm{ROC}(0.2)$, $J$, $\eta$, and $\tau$ are defined
analogously, with their true values denoted by $\mathrm{ROC}_0(0.2)$,
$J_0$, $\eta_0$, and $\tau_0$, respectively. The results are reported
in Table~\ref{tab:qe_point_main_pi90_pi9}. 

\begin{table}[!htt]
\centering
\caption{Point estimation results for $\pi_0=\pi_1=0.90$:
\%RB and MSE, with MSE reported as $100\times\mathrm{MSE}$. Failed fits are excluded from \%RB and MSE calculations.}
\label{tab:qe_point_main_pi90_pi9}
\small
\setlength{\tabcolsep}{2pt}
\resizebox{\textwidth}{!}{%
\begin{tabular}{llc|*{10}{c}|*{10}{c}}
\toprule
& & & \multicolumn{10}{c}{$J_0=0.3$} & \multicolumn{10}{c}{$J_0=0.5$} \\
\cline{4-13}\cline{14-23}
Distribution & $n_0=n_1$ & & \multicolumn{2}{c}{Our} & \multicolumn{2}{c}{NP} & \multicolumn{3}{c}{LBNP} & \multicolumn{3}{c}{LBNP-log} & \multicolumn{2}{c}{Our} & \multicolumn{2}{c}{NP} & \multicolumn{3}{c}{LBNP} & \multicolumn{3}{c}{LBNP-log} \\
\cline{4-5}\cline{6-7}\cline{8-10}\cline{11-13}\cline{14-15}\cline{16-17}\cline{18-20}\cline{21-23}
& & & \%RB & MSE & \%RB & MSE & \%RB & MSE & Fail & \%RB & MSE & Fail & \%RB & MSE & \%RB & MSE & \%RB & MSE & Fail & \%RB & MSE & Fail \\
\midrule
\multirow{15}{*}{\shortstack{Lognormal\\$(\kk=0)$}} & \multirow{5}{*}{100} & $\mathrm{ROC}(0.2)$ & -1.82 & 0.69 & -0.37 & 0.82 & -1.75 & 0.70 & 0 & -1.00 & 0.60 & 0 & -1.53 & 0.62 & -1.02 & 0.79 & -2.21 & 0.68 & \textbf{3} & -1.56 & 0.62 & 0 \\
 &  & AUC & -0.52 & 0.22 & -0.18 & 0.21 & -2.52 & 0.28 & 0 & -0.37 & 0.21 & 0 & -0.57 & 0.15 & -0.05 & 0.17 & -1.93 & 0.21 & \textbf{3} & -0.55 & 0.16 & 0 \\
 &  & $J$ & 0.71 & 0.53 & 18.0 & 0.81 & -3.16 & 0.62 & 0 & 0.58 & 0.49 & 0 & 0.20 & 0.50 & 8.52 & 0.68 & -0.75 & 0.57 & \textbf{3} & -0.32 & 0.49 & 0 \\
 &  & $\eta$ & 0.18 & 1.05 & 4.50 & 1.87 & -10.2 & 1.58 & 0 & -0.02 & 0.55 & 0 & -0.33 & 0.35 & 3.08 & 0.93 & -4.27 & 0.68 & \textbf{3} & 0.08 & 0.31 & 0 \\
 &  & $\tau$ & 0.15 & 1.03 & 3.82 & 1.82 & 8.76 & 1.11 & 0 & 0.29 & 0.54 & 0 & 0.47 & 0.41 & 2.60 & 0.92 & 3.77 & 0.57 & \textbf{3} & -0.29 & 0.32 & 0 \\
\cline{2-23}
 & \multirow{5}{*}{300} & $\mathrm{ROC}(0.2)$ & -0.17 & 0.23 & 0.12 & 0.31 & 0.10 & 0.25 & 0 & -0.17 & 0.21 & 0 & -0.15 & 0.22 & -0.13 & 0.27 & -0.56 & 0.23 & \textbf{1} & -0.30 & 0.22 & 0 \\
 &  & AUC & 0.02 & 0.07 & 0.05 & 0.07 & -1.22 & 0.09 & 0 & -0.04 & 0.07 & 0 & -0.04 & 0.06 & 0.05 & 0.06 & -1.09 & 0.07 & \textbf{1} & -0.11 & 0.06 & 0 \\
 &  & $J$ & 0.80 & 0.17 & 10.4 & 0.29 & -0.07 & 0.20 & 0 & 0.66 & 0.17 & 0 & 0.55 & 0.18 & 4.88 & 0.25 & 0.04 & 0.19 & \textbf{1} & 0.19 & 0.18 & 0 \\
 &  & $\eta$ & 0.31 & 0.26 & 2.46 & 0.99 & -4.90 & 0.73 & 0 & 0.25 & 0.22 & 0 & 0.13 & 0.13 & 1.84 & 0.44 & -1.91 & 0.23 & \textbf{1} & 0.05 & 0.12 & 0 \\
 &  & $\tau$ & 0.06 & 0.27 & 2.31 & 0.99 & 4.86 & 0.61 & 0 & 0.06 & 0.23 & 0 & 0.24 & 0.14 & 1.42 & 0.45 & 1.93 & 0.25 & \textbf{1} & 0.08 & 0.12 & 0 \\
\cline{2-23}
 & \multirow{5}{*}{500} & $\mathrm{ROC}(0.2)$ & -0.21 & 0.14 & 0.00 & 0.19 & 0.13 & 0.16 & \textbf{2} & -0.24 & 0.14 & 0 & -0.16 & 0.14 & 0.02 & 0.16 & -0.30 & 0.14 & \textbf{5} & -0.22 & 0.14 & 0 \\
 &  & AUC & -0.04 & 0.05 & -0.02 & 0.05 & -0.87 & 0.06 & \textbf{2} & -0.08 & 0.05 & 0 & -0.06 & 0.03 & 0.01 & 0.04 & -0.83 & 0.04 & \textbf{5} & -0.10 & 0.04 & 0 \\
 &  & $J$ & 0.27 & 0.11 & 7.39 & 0.18 & 0.25 & 0.13 & \textbf{2} & 0.29 & 0.11 & 0 & 0.21 & 0.11 & 3.58 & 0.15 & 0.08 & 0.12 & \textbf{5} & 0.06 & 0.11 & 0 \\
 &  & $\eta$ & 0.09 & 0.16 & 1.27 & 0.72 & -2.61 & 0.40 & \textbf{2} & 0.13 & 0.13 & 0 & 0.04 & 0.08 & 0.86 & 0.34 & -0.94 & 0.14 & \textbf{5} & 0.08 & 0.07 & 0 \\
 &  & $\tau$ & 0.03 & 0.15 & 2.14 & 0.73 & 2.73 & 0.37 & \textbf{2} & -0.00 & 0.13 & 0 & 0.10 & 0.08 & 1.53 & 0.33 & 0.99 & 0.14 & \textbf{5} & -0.04 & 0.06 & 0 \\
\midrule
\multirow{15}{*}{\shortstack{Weibull\\$(\kk=1/2)$}} & \multirow{5}{*}{100} & $\mathrm{ROC}(0.2)$ & -1.15 & 0.63 & -0.03 & 0.69 & -2.67 & 0.66 & \textbf{6} & -4.36 & 0.68 & 0 & -1.17 & 0.49 & -0.39 & 0.54 & -3.76 & 0.55 & \textbf{19} & -2.53 & 0.58 & 0 \\
 &  & AUC & -0.25 & 0.25 & 0.12 & 0.24 & -2.42 & 0.26 & \textbf{6} & -0.87 & 0.24 & 0 & -0.43 & 0.17 & 0.09 & 0.19 & -3.18 & 0.24 & \textbf{19} & -1.03 & 0.21 & 0 \\
 &  & $J$ & 2.22 & 0.61 & 18.3 & 0.84 & 1.08 & 0.62 & \textbf{6} & -0.97 & 0.56 & 0 & 0.60 & 0.55 & 8.40 & 0.73 & -2.24 & 0.55 & \textbf{19} & -2.56 & 0.58 & 0 \\
 &  & $\eta$ & -1.57 & 1.17 & 8.72 & 1.85 & -8.01 & 1.28 & \textbf{6} & 8.92 & 1.15 & 0 & 0.14 & 0.49 & 4.81 & 0.90 & -4.29 & 0.77 & \textbf{19} & 3.93 & 0.56 & 0 \\
 &  & $\tau$ & 1.94 & 0.93 & 1.13 & 1.52 & 5.90 & 0.88 & \textbf{6} & -6.48 & 1.08 & 0 & 0.25 & 0.33 & 1.28 & 0.63 & 1.94 & 0.33 & \textbf{19} & -4.48 & 0.48 & 0 \\
\cline{2-23}
 & \multirow{5}{*}{300} & $\mathrm{ROC}(0.2)$ & -0.66 & 0.18 & -0.23 & 0.23 & -0.82 & 0.22 & \textbf{3} & -2.48 & 0.21 & 0 & -0.42 & 0.14 & -0.28 & 0.17 & -1.85 & 0.17 & \textbf{11} & -1.41 & 0.16 & 0 \\
 &  & AUC & -0.22 & 0.07 & -0.10 & 0.07 & -1.73 & 0.09 & \textbf{3} & -0.71 & 0.07 & 0 & -0.16 & 0.05 & -0.04 & 0.06 & -2.06 & 0.09 & \textbf{11} & -0.84 & 0.06 & 0 \\
 &  & $J$ & 0.16 & 0.17 & 9.26 & 0.26 & 0.60 & 0.20 & \textbf{3} & -1.32 & 0.17 & 0 & 0.20 & 0.16 & 4.33 & 0.22 & -0.86 & 0.17 & \textbf{11} & -1.95 & 0.17 & 0 \\
 &  & $\eta$ & -0.77 & 0.34 & 3.98 & 0.88 & -4.02 & 0.58 & \textbf{3} & 4.26 & 0.43 & 0 & -0.13 & 0.14 & 2.72 & 0.42 & -2.46 & 0.32 & \textbf{11} & 1.71 & 0.18 & 0 \\
 &  & $\tau$ & 0.59 & 0.27 & 0.87 & 0.73 & 2.99 & 0.45 & \textbf{3} & -3.43 & 0.42 & 0 & 0.22 & 0.11 & 0.48 & 0.35 & 1.36 & 0.20 & \textbf{11} & -2.44 & 0.18 & 0 \\
\cline{2-23}
 & \multirow{5}{*}{500} & $\mathrm{ROC}(0.2)$ & -0.37 & 0.11 & -0.20 & 0.15 & -0.39 & 0.14 & \textbf{2} & -1.69 & 0.13 & 0 & -0.29 & 0.09 & -0.17 & 0.11 & -1.22 & 0.10 & \textbf{13} & -0.99 & 0.10 & 0 \\
 &  & AUC & -0.12 & 0.05 & -0.07 & 0.05 & -1.49 & 0.06 & \textbf{2} & -0.52 & 0.05 & 0 & -0.12 & 0.03 & -0.02 & 0.04 & -1.61 & 0.05 & \textbf{13} & -0.68 & 0.04 & 0 \\
 &  & $J$ & 0.10 & 0.11 & 6.79 & 0.17 & 0.49 & 0.13 & \textbf{2} & -0.89 & 0.11 & 0 & 0.03 & 0.10 & 3.23 & 0.14 & -0.53 & 0.11 & \textbf{13} & -1.53 & 0.11 & 0 \\
 &  & $\eta$ & -0.50 & 0.21 & 3.23 & 0.69 & -3.10 & 0.40 & \textbf{2} & 3.39 & 0.30 & 0 & -0.10 & 0.09 & 1.73 & 0.30 & -1.55 & 0.19 & \textbf{13} & 1.32 & 0.13 & 0 \\
 &  & $\tau$ & 0.38 & 0.14 & 0.43 & 0.58 & 2.32 & 0.30 & \textbf{2} & -2.67 & 0.27 & 0 & 0.10 & 0.06 & 0.58 & 0.24 & 0.87 & 0.14 & \textbf{13} & -1.90 & 0.12 & 0 \\
\midrule
\multirow{15}{*}{\shortstack{Gamma\\$(\kk=1)$}} & \multirow{5}{*}{100} & $\mathrm{ROC}(0.2)$ & -0.67 & 0.60 & -0.03 & 0.69 & -0.89 & 0.59 & 0 & -4.47 & 0.67 & 0 & -1.03 & 0.47 & -0.39 & 0.54 & -1.26 & 0.48 & 0 & -2.54 & 0.57 & 0 \\
 &  & AUC & 0.06 & 0.23 & 0.12 & 0.24 & -0.21 & 0.23 & 0 & -0.86 & 0.24 & 0 & -0.34 & 0.17 & 0.09 & 0.20 & -0.58 & 0.18 & 0 & -1.05 & 0.20 & 0 \\
 &  & $J$ & 2.21 & 0.58 & 18.3 & 0.84 & 1.30 & 0.56 & 0 & -1.22 & 0.55 & 0 & 0.45 & 0.54 & 8.39 & 0.73 & -0.50 & 0.54 & 0 & -2.67 & 0.57 & 0 \\
 &  & $\eta$ & 0.49 & 0.87 & 8.66 & 1.84 & 0.37 & 0.65 & 0 & 9.29 & 1.15 & 0 & 0.43 & 0.44 & 4.83 & 0.90 & -0.02 & 0.40 & 0 & 3.83 & 0.54 & 0 \\
 &  & $\tau$ & 0.52 & 0.70 & 1.17 & 1.52 & 0.25 & 0.41 & 0 & -6.83 & 1.10 & 0 & -0.06 & 0.29 & 1.26 & 0.63 & -0.28 & 0.20 & 0 & -4.47 & 0.47 & 0 \\
\cline{2-23}
 & \multirow{5}{*}{300} & $\mathrm{ROC}(0.2)$ & -0.60 & 0.18 & -0.23 & 0.23 & -0.64 & 0.18 & 0 & -2.53 & 0.21 & 0 & -0.41 & 0.14 & -0.28 & 0.17 & -0.57 & 0.14 & 0 & -1.42 & 0.16 & 0 \\
 &  & AUC & -0.18 & 0.07 & -0.10 & 0.07 & -0.24 & 0.07 & 0 & -0.71 & 0.07 & 0 & -0.16 & 0.05 & -0.04 & 0.06 & -0.33 & 0.05 & 0 & -0.85 & 0.06 & 0 \\
 &  & $J$ & 0.16 & 0.17 & 9.26 & 0.26 & -0.11 & 0.17 & 0 & -1.41 & 0.17 & 0 & 0.17 & 0.16 & 4.33 & 0.22 & -0.38 & 0.16 & 0 & -1.98 & 0.17 & 0 \\
 &  & $\eta$ & -0.51 & 0.30 & 3.98 & 0.88 & -0.24 & 0.23 & 0 & 4.33 & 0.43 & 0 & -0.12 & 0.14 & 2.62 & 0.41 & -0.17 & 0.14 & 0 & 1.66 & 0.18 & 0 \\
 &  & $\tau$ & 0.41 & 0.23 & 0.87 & 0.73 & 0.12 & 0.16 & 0 & -3.51 & 0.42 & 0 & 0.19 & 0.10 & 0.55 & 0.35 & -0.10 & 0.07 & 0 & -2.42 & 0.17 & 0 \\
\cline{2-23}
 & \multirow{5}{*}{500} & $\mathrm{ROC}(0.2)$ & -0.37 & 0.11 & -0.21 & 0.15 & -0.49 & 0.12 & 0 & -1.66 & 0.13 & 0 & -0.30 & 0.09 & -0.17 & 0.11 & -0.44 & 0.09 & 0 & -0.97 & 0.10 & 0 \\
 &  & AUC & -0.12 & 0.05 & -0.07 & 0.05 & -0.15 & 0.04 & 0 & -0.52 & 0.05 & 0 & -0.12 & 0.03 & -0.02 & 0.04 & -0.22 & 0.03 & 0 & -0.68 & 0.04 & 0 \\
 &  & $J$ & 0.09 & 0.11 & 6.79 & 0.17 & -0.18 & 0.11 & 0 & -0.91 & 0.11 & 0 & 0.02 & 0.10 & 3.24 & 0.14 & -0.39 & 0.11 & 0 & -1.49 & 0.11 & 0 \\
 &  & $\eta$ & -0.48 & 0.20 & 3.23 & 0.69 & -0.10 & 0.16 & 0 & 3.21 & 0.29 & 0 & -0.10 & 0.09 & 1.78 & 0.30 & -0.12 & 0.09 & 0 & 1.29 & 0.12 & 0 \\
 &  & $\tau$ & 0.36 & 0.14 & 0.43 & 0.58 & -0.00 & 0.11 & 0 & -2.55 & 0.26 & 0 & 0.09 & 0.06 & 0.55 & 0.24 & -0.14 & 0.05 & 0 & -1.85 & 0.11 & 0 \\
\bottomrule
\end{tabular}%
}
\end{table}

To interpret the performance of LBNP and LBNP-log, it is useful to
consider how their working scales relate to the true density-ratio
structure. LBNP estimates the log-density ratio using B-splines on the
original biomarker scale and is favorably aligned when the true
log-density ratio is linear in $t$. In contrast, LBNP-log applies
B-splines to $\log t$ and is favorably aligned when the true
log-density ratio is linear in $\log t$. Under the BC-DRM, the
lognormal scenario with $\kk=0$ is therefore favorably aligned with
LBNP-log, whereas the gamma scenario with $\kk=1$ is favorably aligned
with LBNP. In the Weibull scenario with $\kk=1/2$, neither working
scale is favorably aligned with the true density-ratio structure.

A clear pattern emerges from Table~\ref{tab:qe_point_main_pi90_pi9}
across $\mathrm{ROC}(0.2)$, AUC, $J$, $\eta$, and $\tau$. When the
working scale is favorably aligned with the true density-ratio
structure, the corresponding LBNP procedure generally performs well,
with Our providing comparable or second-best performance. Specifically,
in the lognormal scenario with $\kk=0$, LBNP-log generally yields the
smallest or nearly the smallest biases and MSEs, while Our performs
similarly in many settings. In the gamma scenario with $\kk=1$, LBNP
generally performs best or nearly best, while Our remains competitive.
The advantage of the favorably aligned LBNP procedure is particularly
apparent for the cutoff-related quantities $\eta$ and $\tau$, whereas
the differences for AUC and $\mathrm{ROC}(0.2)$ are generally smaller.

In the Weibull scenario with $\kk=1/2$, neither LBNP nor LBNP-log has
a favorably aligned working scale. In this case, Our is generally
competitive and often yields the smallest or nearly the smallest
absolute biases and MSEs across the quantities considered, particularly
as the sample size increases. Thus, although a fixed-scale LBNP
procedure can outperform Our when its working scale is favorably
aligned with the underlying density-ratio structure, its performance
is more sensitive to the choice of scale. In contrast, Our estimates
$\kk$ from the data and provides consistently strong performance across
all three scenarios.

The NP estimator exhibits a different pattern. Its estimates of AUC
and $\mathrm{ROC}(0.2)$ are generally competitive, particularly as the
sample size increases. However, it generally performs less favorably for  $J$,
$\eta$, and $\tau$, particularly in terms of MSE. In particular, the relative bias of
$\wh J$ remains appreciable even at the largest sample size
considered. This suggests that the additional structure imposed by the
BC-DRM is particularly beneficial for estimating quantities associated
with the Youden-optimal cutoff.

\begin{figure}[!htbp]
    \centering
    \includegraphics[width=\linewidth]
    {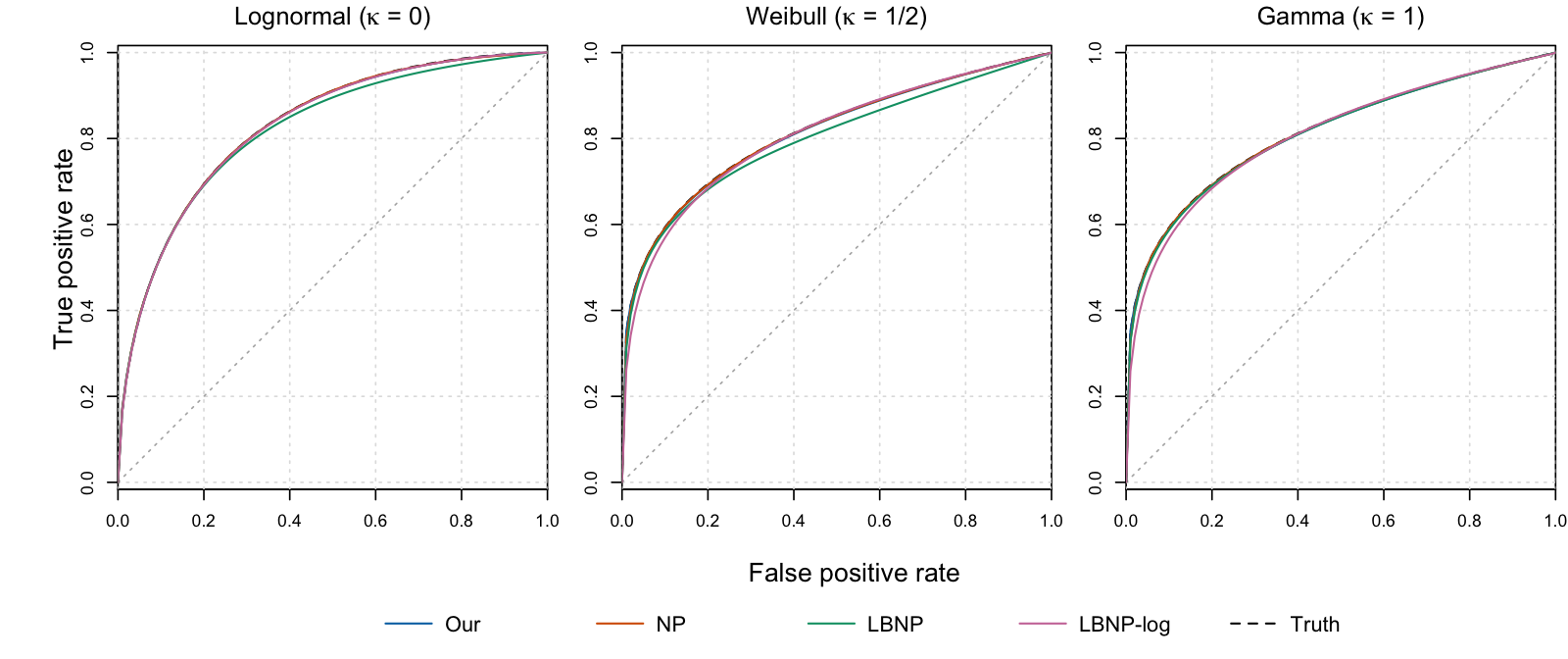}
    \caption{Average estimated ROC curves over 1000 simulation
    replications for $n_0=n_1=300$, $J_0=0.5$, and
    $\pi_0=\pi_1=0.90$.}
    \label{fig:roccurve}
\end{figure}

We next examine estimation of the entire ROC curve.
Figure~\ref{fig:roccurve} displays the average estimated ROC curves over
the 1000 simulation replications, together with the true ROC curves, for
$J_0=0.5$ and $n_0=n_1=300$. The results for $J_0=0.3$ and the other
sample sizes show similar patterns and are provided in Section~5 of the
Supplementary Material. The results are consistent with those for the
scalar quantities. Our produces average estimated ROC curves that
closely track the true curves in all three scenarios. For $\kk=0$,
LBNP-log also closely tracks the true ROC curve, whereas LBNP exhibits
some systematic deviation. For $\kk=1$, LBNP closely tracks the true
ROC curve, while LBNP-log shows some systematic deviation. For
$\kk=1/2$, where neither fixed working scale is favorably aligned with
the true density-ratio structure, both LBNP and LBNP-log exhibit
systematic deviations from the true ROC curve, but in different
regions: LBNP-log tends to underestimate the ROC curve in the lower
false-positive-rate region, whereas LBNP tends to underestimate it in
the higher false-positive-rate region. In contrast, Our closely tracks
the true ROC curve over the entire range.

Finally, the fitting-failure counts in
Table~\ref{tab:qe_point_main_pi90_pi9} reveal differences in numerical
stability. LBNP exhibits occasional fitting failures in the lognormal
scenario and more frequent failures in the Weibull scenario,
particularly when $J_0=0.5$, whereas no fitting failures are observed
for Our, NP, or LBNP-log. Taken together, the results reveal a clear
trade-off. When the working scale is favorably aligned with the true
density-ratio structure, the corresponding LBNP procedure can perform
particularly well, with Our generally remaining competitive. When
neither fixed working scale is favorably aligned, Our generally
provides more robust performance. By estimating $\kk$ from the data,
Our therefore provides accurate and numerically stable estimation
across different transformation structures without requiring prior
knowledge of the appropriate transformation scale.


\subsection{Evaluation of confidence intervals}

We evaluate the performance of the percentile bootstrap CIs using
coverage probability (CP) and average length (AL). Let
$\mathcal{I}_{\mathrm{AUC}}^{(i)}$ denote the CI for AUC obtained from
the $i$th simulation replicate, and let
$L_{\mathrm{AUC}}^{(i)}$ denote its length. We define
\[
\mathrm{CP}(\mathrm{AUC})
=
\frac{1}{N}
\sum_{i=1}^{N}
I\left\{
\mathrm{AUC}_0\in
\mathcal{I}_{\mathrm{AUC}}^{(i)}
\right\},
\qquad
\mathrm{AL}(\mathrm{AUC})
=
\frac{1}{N}
\sum_{i=1}^{N}
L_{\mathrm{AUC}}^{(i)},
\]
where $\mathrm{AUC}_0$ denotes the true AUC. The corresponding criteria
for $\mathrm{ROC}(0.2)$, $J$, $\eta$, and $\tau$ are defined
analogously. The results are summarized in
Table~\ref{tab:qe_coverage_main_pi90_pi9}.

\begin{table}[!htt]
\centering
\caption{Coverage probability and average length of the 95\% percentile
bootstrap CIs for $\pi_0=\pi_1=0.90$.}
\label{tab:qe_coverage_main_pi90_pi9}
\small
\setlength{\tabcolsep}{2pt}
\resizebox{\textwidth}{!}{%
\begin{tabular}{llc|*{8}{c}|*{8}{c}}
\toprule
& & & \multicolumn{8}{c}{$J_0=0.3$} & \multicolumn{8}{c}{$J_0=0.5$} \\
\cline{4-11}\cline{12-19}
Distribution & $n_0=n_1$ & & \multicolumn{2}{c}{Our} & \multicolumn{2}{c}{NP} & \multicolumn{2}{c}{LBNP} & \multicolumn{2}{c}{LBNP-log} & \multicolumn{2}{c}{Our} & \multicolumn{2}{c}{NP} & \multicolumn{2}{c}{LBNP} & \multicolumn{2}{c}{LBNP-log} \\
\cline{4-5}\cline{6-7}\cline{8-9}\cline{10-11}\cline{12-13}\cline{14-15}\cline{16-17}\cline{18-19}
& & & CP (\%) & AL & CP (\%) & AL & CP (\%) & AL & CP (\%) & AL & CP (\%) & AL & CP (\%) & AL & CP (\%) & AL & CP (\%) & AL \\
\midrule
\multirow{15}{*}{\shortstack{Lognormal\\$(\kk=0)$}} & \multirow{5}{*}{100} & $\mathrm{ROC}(0.2)$ & 94.5 & 0.33 & 96.9 & 0.37 & 95.5 & 0.34 & 95.7 & 0.32 & 96.0 & 0.32 & 97.0 & 0.35 & 95.6 & 0.33 & 96.2 & 0.32 \\
 &  & AUC & 95.9 & 0.19 & 95.1 & 0.19 & 95.1 & 0.19 & 95.2 & 0.18 & 96.6 & 0.16 & 95.2 & 0.16 & 95.5 & 0.16 & 96.1 & 0.16 \\
 &  & $J$ & 95.5 & 0.30 & 80.5 & 0.28 & 96.4 & 0.30 & 95.7 & 0.28 & 95.7 & 0.29 & 85.3 & 0.28 & 96.8 & 0.29 & 96.7 & 0.28 \\
 &  & $\eta$ & 96.4 & 0.41 & 97.1 & 0.47 & 98.0 & 0.44 & 98.2 & 0.40 & 96.1 & 0.25 & 95.6 & 0.34 & 98.6 & 0.31 & 97.3 & 0.28 \\
 &  & $\tau$ & 95.6 & 0.41 & 97.3 & 0.47 & 94.9 & 0.40 & 98.4 & 0.40 & 96.4 & 0.27 & 97.9 & 0.34 & 97.6 & 0.30 & 97.9 & 0.27 \\
\cline{2-19}
 & \multirow{5}{*}{300} & $\mathrm{ROC}(0.2)$ & 94.2 & 0.19 & 95.2 & 0.22 & 94.6 & 0.20 & 94.6 & 0.19 & 94.5 & 0.18 & 95.9 & 0.20 & 94.9 & 0.19 & 94.2 & 0.19 \\
 &  & AUC & 94.9 & 0.11 & 95.1 & 0.11 & 95.2 & 0.11 & 95.2 & 0.11 & 94.6 & 0.09 & 94.4 & 0.09 & 94.4 & 0.09 & 95.3 & 0.09 \\
 &  & $J$ & 94.5 & 0.17 & 82.7 & 0.17 & 95.5 & 0.17 & 95.5 & 0.17 & 93.9 & 0.16 & 85.6 & 0.17 & 95.5 & 0.17 & 94.7 & 0.16 \\
 &  & $\eta$ & 95.3 & 0.22 & 98.3 & 0.36 & 96.9 & 0.30 & 97.3 & 0.27 & 95.2 & 0.14 & 98.2 & 0.25 & 98.4 & 0.20 & 97.7 & 0.17 \\
 &  & $\tau$ & 94.9 & 0.22 & 98.1 & 0.36 & 95.2 & 0.29 & 97.2 & 0.27 & 94.3 & 0.15 & 96.7 & 0.25 & 95.8 & 0.20 & 97.2 & 0.17 \\
\cline{2-19}
 & \multirow{5}{*}{500} & $\mathrm{ROC}(0.2)$ & 94.1 & 0.14 & 94.7 & 0.17 & 94.2 & 0.16 & 94.9 & 0.15 & 95.2 & 0.14 & 95.7 & 0.16 & 94.8 & 0.15 & 95.7 & 0.15 \\
 &  & AUC & 94.8 & 0.08 & 93.9 & 0.08 & 93.9 & 0.08 & 94.3 & 0.08 & 95.8 & 0.07 & 95.0 & 0.07 & 93.2 & 0.07 & 94.8 & 0.07 \\
 &  & $J$ & 94.6 & 0.13 & 83.7 & 0.14 & 94.5 & 0.14 & 94.2 & 0.13 & 95.9 & 0.13 & 86.4 & 0.13 & 95.5 & 0.13 & 96.1 & 0.13 \\
 &  & $\eta$ & 93.6 & 0.16 & 98.6 & 0.31 & 97.3 & 0.25 & 97.6 & 0.22 & 94.3 & 0.11 & 98.0 & 0.21 & 96.7 & 0.15 & 96.8 & 0.14 \\
 &  & $\tau$ & 94.4 & 0.16 & 98.2 & 0.31 & 95.9 & 0.24 & 98.4 & 0.22 & 94.6 & 0.11 & 98.0 & 0.21 & 96.3 & 0.15 & 97.2 & 0.14 \\
\midrule
\multirow{15}{*}{\shortstack{Weibull\\$(\kk=1/2)$}} & \multirow{5}{*}{100} & $\mathrm{ROC}(0.2)$ & 94.5 & 0.31 & 94.9 & 0.32 & 94.5 & 0.30 & 93.7 & 0.30 & 94.6 & 0.28 & 93.9 & 0.28 & 94.0 & 0.27 & 93.3 & 0.28 \\
 &  & AUC & 94.5 & 0.19 & 93.5 & 0.19 & 94.5 & 0.18 & 93.8 & 0.18 & 94.3 & 0.16 & 93.7 & 0.17 & 91.4 & 0.16 & 93.7 & 0.16 \\
 &  & $J$ & 94.1 & 0.30 & 79.6 & 0.28 & 95.3 & 0.29 & 94.0 & 0.28 & 94.3 & 0.29 & 85.6 & 0.27 & 95.3 & 0.28 & 94.5 & 0.27 \\
 &  & $\eta$ & 96.3 & 0.45 & 96.0 & 0.47 & 98.1 & 0.41 & 96.1 & 0.43 & 95.0 & 0.29 & 93.6 & 0.34 & 98.0 & 0.32 & 95.6 & 0.30 \\
 &  & $\tau$ & 94.2 & 0.39 & 97.5 & 0.43 & 96.4 & 0.34 & 98.0 & 0.40 & 96.2 & 0.22 & 97.5 & 0.29 & 98.5 & 0.24 & 98.0 & 0.26 \\
\cline{2-19}
 & \multirow{5}{*}{300} & $\mathrm{ROC}(0.2)$ & 96.7 & 0.17 & 96.0 & 0.19 & 95.5 & 0.18 & 95.4 & 0.18 & 95.9 & 0.15 & 95.6 & 0.17 & 95.3 & 0.16 & 95.2 & 0.16 \\
 &  & AUC & 96.3 & 0.11 & 95.1 & 0.11 & 95.6 & 0.11 & 95.7 & 0.11 & 95.5 & 0.09 & 95.0 & 0.10 & 91.5 & 0.09 & 95.1 & 0.09 \\
 &  & $J$ & 96.3 & 0.17 & 83.9 & 0.17 & 96.1 & 0.17 & 96.6 & 0.17 & 95.7 & 0.16 & 87.8 & 0.17 & 96.4 & 0.16 & 96.0 & 0.16 \\
 &  & $\eta$ & 95.6 & 0.24 & 95.8 & 0.34 & 98.3 & 0.29 & 95.9 & 0.28 & 96.0 & 0.16 & 96.5 & 0.24 & 97.5 & 0.21 & 95.8 & 0.19 \\
 &  & $\tau$ & 95.5 & 0.21 & 97.3 & 0.32 & 96.6 & 0.25 & 96.5 & 0.26 & 94.9 & 0.13 & 98.0 & 0.22 & 96.4 & 0.17 & 96.1 & 0.16 \\
\cline{2-19}
 & \multirow{5}{*}{500} & $\mathrm{ROC}(0.2)$ & 95.0 & 0.13 & 95.1 & 0.15 & 95.2 & 0.14 & 94.2 & 0.14 & 95.4 & 0.12 & 94.2 & 0.13 & 94.6 & 0.12 & 94.2 & 0.12 \\
 &  & AUC & 94.2 & 0.08 & 93.7 & 0.08 & 93.8 & 0.08 & 93.9 & 0.08 & 95.1 & 0.07 & 93.9 & 0.07 & 90.3 & 0.07 & 93.7 & 0.07 \\
 &  & $J$ & 94.5 & 0.13 & 83.8 & 0.13 & 95.1 & 0.14 & 94.9 & 0.13 & 94.2 & 0.12 & 87.9 & 0.13 & 95.3 & 0.13 & 94.2 & 0.12 \\
 &  & $\eta$ & 94.8 & 0.18 & 96.7 & 0.29 & 97.1 & 0.24 & 94.6 & 0.23 & 94.2 & 0.12 & 95.8 & 0.20 & 95.6 & 0.17 & 94.8 & 0.15 \\
 &  & $\tau$ & 95.1 & 0.16 & 98.9 & 0.28 & 96.1 & 0.22 & 97.7 & 0.22 & 95.5 & 0.10 & 98.0 & 0.18 & 96.5 & 0.14 & 96.0 & 0.13 \\
\midrule
\multirow{15}{*}{\shortstack{Gamma\\$(\kk=1)$}} & \multirow{5}{*}{100} & $\mathrm{ROC}(0.2)$ & 93.8 & 0.30 & 94.6 & 0.32 & 93.7 & 0.30 & 93.1 & 0.29 & 94.0 & 0.27 & 93.9 & 0.28 & 93.9 & 0.27 & 93.1 & 0.27 \\
 &  & AUC & 94.1 & 0.18 & 93.3 & 0.19 & 93.9 & 0.18 & 93.8 & 0.18 & 94.1 & 0.16 & 93.4 & 0.17 & 94.2 & 0.16 & 93.1 & 0.16 \\
 &  & $J$ & 93.6 & 0.29 & 80.5 & 0.28 & 94.2 & 0.28 & 94.5 & 0.28 & 94.5 & 0.28 & 85.8 & 0.27 & 95.1 & 0.27 & 94.4 & 0.27 \\
 &  & $\eta$ & 96.4 & 0.38 & 96.0 & 0.47 & 97.8 & 0.40 & 95.5 & 0.41 & 94.7 & 0.27 & 93.9 & 0.34 & 96.8 & 0.30 & 95.3 & 0.29 \\
 &  & $\tau$ & 95.1 & 0.34 & 98.1 & 0.43 & 97.9 & 0.34 & 97.9 & 0.38 & 95.9 & 0.21 & 96.8 & 0.29 & 98.3 & 0.24 & 97.6 & 0.24 \\
\cline{2-19}
 & \multirow{5}{*}{300} & $\mathrm{ROC}(0.2)$ & 96.4 & 0.17 & 95.9 & 0.19 & 96.5 & 0.17 & 95.0 & 0.17 & 95.9 & 0.15 & 95.5 & 0.17 & 96.3 & 0.15 & 94.7 & 0.15 \\
 &  & AUC & 95.8 & 0.11 & 95.5 & 0.11 & 95.8 & 0.11 & 95.4 & 0.10 & 95.9 & 0.09 & 94.8 & 0.10 & 95.7 & 0.09 & 94.9 & 0.09 \\
 &  & $J$ & 96.3 & 0.17 & 83.4 & 0.17 & 96.9 & 0.17 & 96.7 & 0.16 & 95.8 & 0.16 & 88.4 & 0.17 & 95.9 & 0.16 & 95.7 & 0.16 \\
 &  & $\eta$ & 96.1 & 0.21 & 95.7 & 0.34 & 98.6 & 0.27 & 95.4 & 0.28 & 96.0 & 0.15 & 96.4 & 0.24 & 96.7 & 0.18 & 95.6 & 0.18 \\
 &  & $\tau$ & 95.5 & 0.18 & 98.1 & 0.32 & 98.7 & 0.24 & 96.4 & 0.26 & 94.8 & 0.12 & 98.0 & 0.22 & 98.1 & 0.15 & 95.9 & 0.15 \\
\cline{2-19}
 & \multirow{5}{*}{500} & $\mathrm{ROC}(0.2)$ & 94.4 & 0.13 & 95.5 & 0.15 & 94.3 & 0.13 & 94.4 & 0.13 & 94.7 & 0.11 & 94.6 & 0.13 & 94.7 & 0.12 & 93.8 & 0.12 \\
 &  & AUC & 94.1 & 0.08 & 93.7 & 0.08 & 94.5 & 0.08 & 93.7 & 0.08 & 94.9 & 0.07 & 93.9 & 0.07 & 94.9 & 0.07 & 93.8 & 0.07 \\
 &  & $J$ & 94.4 & 0.13 & 84.0 & 0.13 & 94.6 & 0.13 & 94.2 & 0.13 & 95.0 & 0.12 & 87.7 & 0.13 & 94.8 & 0.13 & 94.3 & 0.12 \\
 &  & $\eta$ & 94.7 & 0.17 & 97.0 & 0.29 & 96.4 & 0.22 & 94.3 & 0.23 & 94.4 & 0.12 & 96.1 & 0.20 & 96.2 & 0.15 & 94.1 & 0.14 \\
 &  & $\tau$ & 95.6 & 0.14 & 98.7 & 0.28 & 98.2 & 0.20 & 96.9 & 0.21 & 94.8 & 0.10 & 97.8 & 0.18 & 97.8 & 0.12 & 95.3 & 0.12 \\
\bottomrule
\end{tabular}%
}
\end{table}

The behavior of LBNP and LBNP-log is broadly consistent with the
point-estimation results. When the working scale is favorably aligned
with the true density-ratio structure, the corresponding procedure
generally provides good coverage with competitive interval lengths.
When the working scale is less favorable, some deterioration in
coverage can occur. For example, LBNP exhibits noticeable AUC
undercoverage in the Weibull setting with $\kk=1/2$ and $J_0=0.5$,
consistent with the systematic bias observed in its point estimates.
In contrast, Our maintains coverage close to the nominal level without
requiring the transformation scale to be specified in advance.

In terms of interval length, the four methods are generally comparable
for AUC. For $\mathrm{ROC}(0.2)$, NP generally produces the longest
intervals, whereas Our, LBNP, and LBNP-log yield shorter and broadly
comparable intervals. More pronounced differences are observed for
$\eta$ and $\tau$, for which NP again generally produces the longest
intervals. Our  generally  yields substantially shorter intervals while maintaining
coverage close to the nominal level, and its interval lengths are
generally shorter than, or comparable to, those of LBNP and LBNP-log.
For $J$, the interval lengths of the four methods are generally similar;
however, NP exhibits substantial undercoverage, whereas Our, LBNP, and
LBNP-log generally maintain coverage close to the nominal level.
Overall, Our provides a favorable balance between coverage accuracy and
interval length across the quantities and distributional settings
considered.

\section{Real data analysis\label{bcdrm.sec6}}

We illustrate the proposed method using data from a malaria study conducted
in the Kilombero district of Tanzania \citep{Kitua1996,qin2005semiparametric}.
Malaria transmission in this region is highly seasonal, with substantially
higher prevalence during the wet season than during the dry season. The
study recorded parasite density, measured as the estimated number of
parasites per $\mu$L of blood, for children examined during the wet and
dry seasons. The dry-season sample has substantially lower malaria
prevalence, whereas the wet-season sample contains a mixture of individuals with and without malaria.
The seasonal grouping therefore plays the role of an imperfect
reference standard, with the dry- and wet-season samples having different
underlying malaria prevalences.

We restrict the analysis to individuals with positive recorded parasite
density, yielding 81 children examined during the dry season and 211
children examined during the wet season. Let $R=0$ and $R=1$ denote the
nominal disease statuses corresponding to the dry- and wet-season samples,
respectively. Following the previous analysis of these data
\citep{Sun2025Likelihood}, we take
\[
\pi_0=\Pr(D=0\mid R=0)=1
\quad\text{and}\quad
\pi_1=\Pr(D=1\mid R=1)=0.677.
\]
Thus, individuals in the dry-season sample are treated as truly
nonmalaria, whereas 67.7\% of those in the wet-season sample are assumed
to have malaria.

The recorded parasite densities range from 1 to 399,952.1 parasites per
$\mu$L of blood. For numerical convenience, we rescale the parasite
density by dividing all observations by $100{,}000$ and use the rescaled
values as the biomarker $T$ in the subsequent analysis. This positive
linear rescaling does not affect the ROC curve or its associated summary
measures. Moreover, both the proposed BC-DRM and LBNP are theoretically
invariant to such rescaling, although their numerical implementations may
be affected by the scale of the observations. We therefore use the
rescaled data to improve numerical conditioning.

We first assess the adequacy of the proposed BC-DRM~\eqref{def.bcdrm}
using the goodness-of-fit procedure developed in Section~\ref{bcdrm.main.sec3.3}. Based on
$M=500$ bootstrap samples, the test yields a $p$-value of 0.599,
providing no evidence against the BC-DRM.

\begin{figure}[!htt]
    \centering
    \includegraphics[width=1\linewidth]{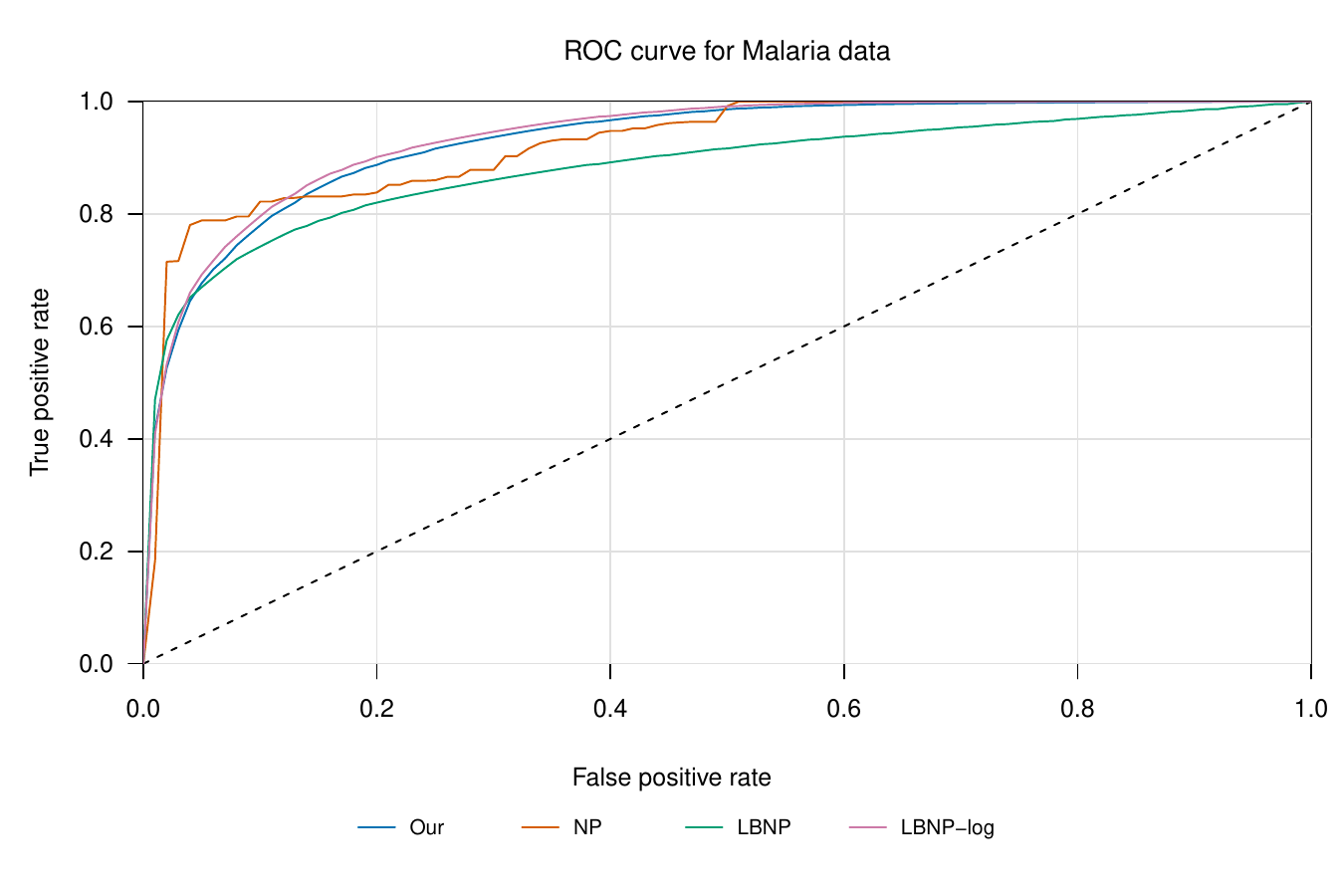}
    \caption{Estimated ROC curves for the malaria data.}
    \label{fig:ROC_real_data}
\end{figure}

We next apply the proposed method, together with the NP, LBNP, and
LBNP-log methods considered in the simulation study, to the malaria
data. Table~\ref{tab:malaria-div100000} reports the estimates of
$\mathrm{ROC}(0.2)$, AUC, $J$, $\eta$, and $\tau$, together with their
95\% percentile bootstrap CIs based on $M=500$ bootstrap samples,
while Figure~\ref{fig:ROC_real_data} displays the estimated ROC curves
from the four methods. For the proposed method, the estimated Box--Cox
transformation parameter is
\(
\widehat{\kk}=0.1821,
\)
with a 95\% percentile bootstrap CI of $(-0.7981,\,0.5451)$. The point
estimate is closer to $\kk=0$, corresponding to a logarithmic
transformation, than to $\kk=1$, corresponding to the identity
transformation. Moreover, the CI contains $\kk=0$ but excludes
$\kk=1$, providing support for a transformation closer to the
logarithmic scale than to the identity scale.

\begin{table}[!htt]
\centering
\caption{Point estimates for the malaria data, with 95\% percentile
bootstrap CIs in parentheses.}
\label{tab:malaria-div100000}
\resizebox{\textwidth}{!}{%
\begin{tabular}{lccccc}
\toprule
Method
& $\mathrm{ROC}(0.2)$
& AUC
& $J$
& $\eta$
& $\tau$ \\
\midrule
Our
& 0.887 (0.763, 1.000)
& 0.931 (0.866, 0.993)
& 0.697 (0.570, 0.911)
& 0.853 (0.755, 0.966)
& 0.844 (0.779, 0.948) \\

NP
& 0.838 (0.734, 0.968)
& 0.932 (0.849, 1.004)
& 0.743 (0.646, 0.871)
& 0.780 (0.682, 0.948)
& 0.963 (0.845, 1.000) \\

LBNP
& 0.820 (0.749, 0.971)
& 0.885 (0.841, 0.972)
& 0.644 (0.576, 0.854)
& 0.766 (0.689, 0.918)
& 0.878 (0.855, 0.964) \\

LBNP-log
& 0.901 (0.723, 0.964)
& 0.937 (0.836, 0.965)
& 0.712 (0.580, 0.835)
& 0.862 (0.664, 0.910)
& 0.850 (0.815, 0.971) \\
\bottomrule
\end{tabular}%
}

\par\smallskip
\begin{minipage}{0.98\linewidth}
\footnotesize
\textit{Note:} CIs are based on successful bootstrap fits with valid
ROC curves. The numbers of successful bootstrap fits out of 500 are
500 for Our, NP, and LBNP-log, and 482 for LBNP. Failed bootstrap fits are excluded from CI calculations.
\end{minipage}
\end{table}

The point estimates in Table~\ref{tab:malaria-div100000} are consistent
with the estimated transformation parameter. Across all five ROC summary
measures, the estimates from LBNP-log are close to those from the proposed
method, whereas the corresponding LBNP estimates show larger differences.
This agrees with the estimate $\widehat{\kk}=0.1821$, which favors a
transformation closer to the logarithmic scale than to the identity scale.

The NP method gives an AUC estimate of 0.932, which is nearly identical
to the estimate of 0.931 from the proposed method. More noticeable
differences are observed for the other ROC summary measures, particularly
for the Youden-optimal sensitivity and specificity. Thus, close agreement
in AUC does not necessarily translate into similar estimates of the ROC
curve at a particular false-positive rate or of the Youden-optimal
operating characteristics.

It is also worth noting that the upper endpoint of the 95\% percentile
bootstrap CI for AUC obtained by NP is slightly greater than one. The NP
estimator of AUC is obtained by inverting the relationship between the
AUC based on the nominal disease status and the true AUC. The resulting
estimator is not constrained to $[0,1]$, so some bootstrap estimates can
exceed one, resulting in an upper confidence limit above one.

Turning to estimation of the entire ROC curve,
Figure~\ref{fig:ROC_real_data} provides further insight into these
comparisons. The ROC curves estimated by the proposed method and
LBNP-log are close over most of the false-positive-rate range, whereas
the LBNP curve shows a more noticeable departure. The NP curve also
differs locally from the proposed curve despite their nearly identical
AUC estimates. These results further illustrate that similar AUC
estimates can mask differences in other aspects of ROC performance.

\section{Discussion\label{bcdrm.sec7}}

In this paper, we proposed a semiparametric framework for ROC analysis
in the presence of an imperfect reference standard. The proposed BC-DRM
combines the flexibility of the density ratio model with the Box--Cox
transformation, allowing the transformation parameter to be estimated
from the data rather than requiring a prespecified transformation scale.
We developed empirical-likelihood-based estimation for the ROC curve and
its associated summary measures, established the corresponding
asymptotic properties, and proposed bootstrap procedures for statistical
inference and goodness-of-fit assessment. An EM algorithm was developed
for numerical implementation. Simulation studies demonstrated that the proposed method provides accurate and numerically stable estimation and inference across different transformation structures, and the malaria data analysis illustrated its practical applicability.

One limitation of the current framework is that $\pi_0$ and $\pi_1$ in \eqref{pi.def} are assumed to be known. This assumption
plays an important role in the identifiability of the model, and allowing
both quantities to be unknown may lead to an identifiability issue under
the current framework. Developing methods that incorporate uncertainty
in $\pi_0$ and $\pi_1$, possibly by introducing additional information
or assumptions to ensure identifiability, is an important direction for
future research.

Another direction is to extend the proposed framework beyond binary
disease status to settings involving multiple classes. Such an extension
would require suitable generalizations of the BC-DRM, together with the
development of corresponding ROC-type measures and statistical inference.

Finally, we have focused on Youden's index as the criterion for selecting
an optimal cutoff point. Other criteria may be preferable in
different applications, such as selecting the point on the ROC curve
closest to $(0,1)$ or the cutoff at which sensitivity and specificity
are equal \citep{sande2021statistical}. It would be interesting to extend the proposed framework to
estimation and inference for optimal cutoff points and their associated
operating characteristics under these alternative criteria.

\section*{Acknowledgements}

This work originated in part from Yi Chang's STAT 900 research project at the University of Waterloo.
The authors thank Professors Mu Zhu, Yingli Qin, and Peijun Sang for their
helpful comments and suggestions, which substantially improved the paper. Qinglong Tian's research is supported by the Natural Sciences
and Engineering Research Council of Canada (NSERC; RGPIN-2023-03479).
Pengfei Li's research is supported in part by NSERC (RGPIN-2026-04406)
and the Faculty of Mathematics Research Chair at the University of Waterloo.

\bibliographystyle{chicago}
\bibliography{ref}
\end{document}